\documentclass[cits]{PoS}

\title{Flaring Activity of Sgr A*: Expanding Hot Blobs}
\ShortTitle{Sgr A*}

\author{\speaker{F. Yusef-Zadeh$^1$}, M. Wardle$^2$, D. A. Roberts$^3$, 
C. O. Heinke$^4$, C. D. Dowell$^5$, W. D. Cotton$^6$, G. C. Bower$^7$ \& 
F. K. Baganoff$^8$\\

$^1$ Dept of Physics and Astronomy, 
Northwestern University, Evanston, Il. 60208\\
        E-mail: \email{zadeh@northwestern.edu}\\
$^2$Department of Physics, Macquarie University, Sydney NSW 2109,
Australia\\
        E-mail: \email{wardle@ics.mq.edu.au}\\
      $^3$Adler Planetarium and Astronomy Museum, 1300
South Lake Shore Drive, Chicago, IL 60605 \\
        E-mail: \email{doug-roberts@northwestern.edu}\\
$^4$Department of Physics and Astronomy,
Northwestern University, Evanston, Il. 60208\\
        E-mail: \email{cheinke@northwestern.edu}\\
       $^5$Cal Tech, Jet Propulsion Laboratory, Pasadena, CA 91109 \\
        E-mail: \email{cdd@submm.caltech.edu}\\
       $^6$National Radio Astronomy Observatory, 520 Edgemont Road, 
Charlottesville, VA 22903\\
        E-mail: \email{bcotton@NRAO.EDU}\\
       $^7$Radio Astronomy Lab, 601 Campbell Hall,
University of California, Berkeley, CA 94720\\
        E-mail: \email{gbower@astro.berkeley.edu}\\
        $^8$Kavli Institute for Astrophysics and Space Research, MIT, 
Cambridge, 
MA 02139-4307\\
        E-mail: \email{fkb@space.mit.edu}}

\abstract{
Sgr A* is considered to be a massive black hole at the Galactic center and
is known to be variable in radio, millimeter, near-IR and X-rays. Recent
multi-wavelength observing campaigns show a simultaneous X-ray and near-IR
flare, as well as sub-millimeter and near-IR flares from Sgr A*. The flare
activity is thought to be arising from the innermost region of Sgr A*. 
We have recently argued that the duration of flares in near-IR and
submillimeter wavelengths
implies that the burst of emission expands and cools on a dynamical time
scale before the flares leave Sgr A*. The  detection of radio flares 
 with a time delay in the range of  20 and 40 minutes between 7 and 12mm peak emission  implies
adiabatic expansion of a uniform, spherical hot  blob due to flare
activity. We suspect that this simple outflow picture shows some of 
the characteristics that are known to take place in 
microquasars, thus we may learn much from comparative 
study of Sgr A* and its environment vs. microquasars.}

\FullConference{VI Microquasar Workshop: Microquasars and Beyond\\
		September 18-22 2006\\
		Societ\`a del Casino, Como, Italy}

\def\ee #1 {\times 10^{#1}}          
\def\ut #1 #2 { \, \mathrm{#1}^{#2}} 
\def\u #1 { \, \mathrm{#1}}          
\def\msol{\hbox{$\hbox{M}_\odot$}}

\def\msol       {\hbox{$\hbox{M}_\odot$}}

\begin{document}

\section{Background}

More than three decades have elapsed since the discovery of Sgr A* 
(Balick \& Brown 1974), and during most of 
this time the source remained undetected outside the radio band. Recent observations have provided compelling 
evidence that the compact nonthermal radio source Sgr A* can be clearly 
identified with the massive black hole 
at the center of the Galaxy.  A major breakthrough in our understanding of the nature of this source came from 
stellar orbital measurements showing a mass 3-4 $\times 10^6$ \msol 
coincident within 45 AU of Sgr A* (e.g., 
Sch\"odel et al. 2003; Ghez et al. 2003).  This dark, massive object has been uniquely identified with the 
radio source Sgr A* through measurements of the proper motion of the radio source, which show that Sgr A* must 
contain $>4\times10^{5}$ \msol and does not orbit another massive object (Reid \& Brunthaler 2004).  A 
``sub-millimeter bump'' was detected in the broad-band spectrum of Sgr 
A* (Zylka et al. 1995; Falcke et al. 
1998); the peak sub-millimeter frequency is thought to be the dividing line between optically thin emission at 
higher frequencies, and optically thick emission at lower frequencies. A number of additional observations 
such as the time variability and polarization (Bower et al. 2005; Marrone et al. 2006) analysis are providing 
additional opportunities to study the innermost regions, within just a f
ew Schwarzschild radii of the black 
hole event horizon, a region currently unaccessible via high spatial resolution imaging. Here we focus only on 
the recent time variability studies of Sgr A*.


The {\it Chandra} X-ray  Observatory 
detected the quiescent X-ray counterpart of Sgr A$^*$, followed a year later by 
the discovery of strong X-ray flares by {\it Chandra} (Baganoff et al.\ 2001) and then 
XMM-Newton (Goldwurm et al. 2003). During a 
flare event, the X-ray flux of Sgr A$^*$ increases by 1--2 orders of 
magnitude.  Using XMM-Newton, Porquet et al. (2003) detected the most 
powerful X-ray flare seen to date, which also showed the softest spectrum.  
The typical time scale for X-ray flares is about an hour. 
More recently, the 
long-sought near-IR counterpart to Sgr A$^*$ was discovered when Genzel et 
al. (2003) and Ghez et al. (2004) detected quiescent emission and a number of 
flares in the H, K$_s$ and L' bands from Sgr A$^*$.
Typical IR flares last 
about 40 minutes, with the flux increasing by a factor of about 5.  
Variability has also been reported at millimeter and centimeter 
wavelengths (Zhao et al.\ 2003; Miyazaki et al.\ 2004). 
The amplitude is small, however, in 
comparison to that of  X-ray and near-IR wavelengths, 
and the variability time scale appears to range between days and weeks. 

Although numerous observations have detected Sgr A* in its flaring state, there are limited simultaneous 
observations to measure the correlation in different wavelength bands and within radio bands 
(e.g., Eckart et 
al.\ 2006). On the theoretical side numerous models have also been proposed. The discussion of 
these models are beyond the scope of this contribution paper.  
Highlights of our multi-wavelength observations 
are presented here (Yusef-Zadeh et al. 2006a,b), 
as the following issues are discussed:  Is the time variability of Sgr A* at different wavelengths correlated?  
And if so, what will correlated variability measurements tell us about the emission mechanisms and the properties of the
accretion flow within a few Schwarzschild radii of the black hole?


\section {Correlation Study}
We begin  with the correlation study of X-ray and 
near-IR 
flares, followed by the correlation of near-IR and sub-millimeter 
wavelengths. 
Finally, we present the correlation of flare emission in
submillimeter and radio bands. The 
discussion below should provide some insights on the physical 
processes that may  run parallel  in both micro-quasars and 
Sgr A*.

\begin{figure}
\includegraphics[width=0.5\textwidth]{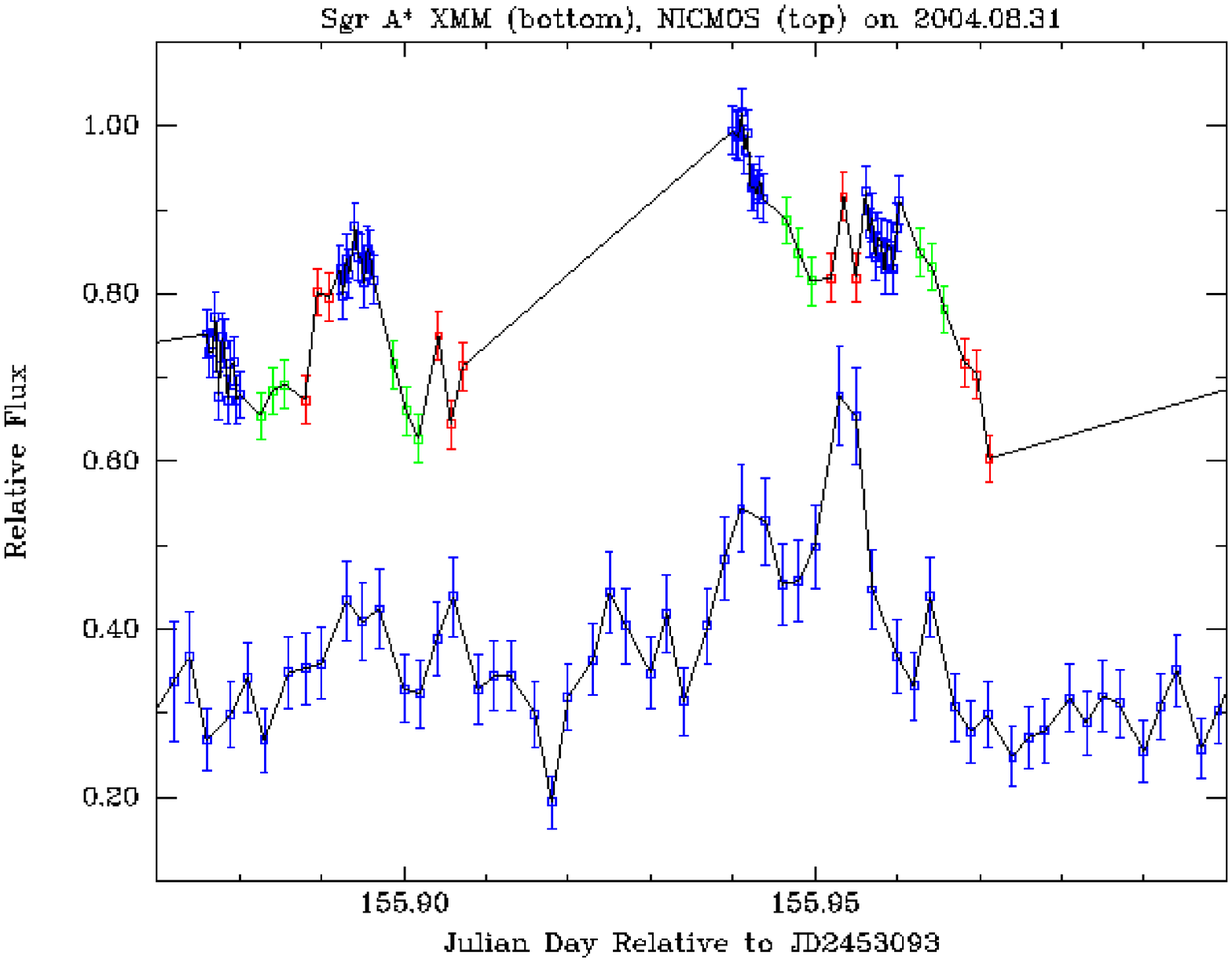}
\includegraphics[width=0.4\textwidth,angle=0]{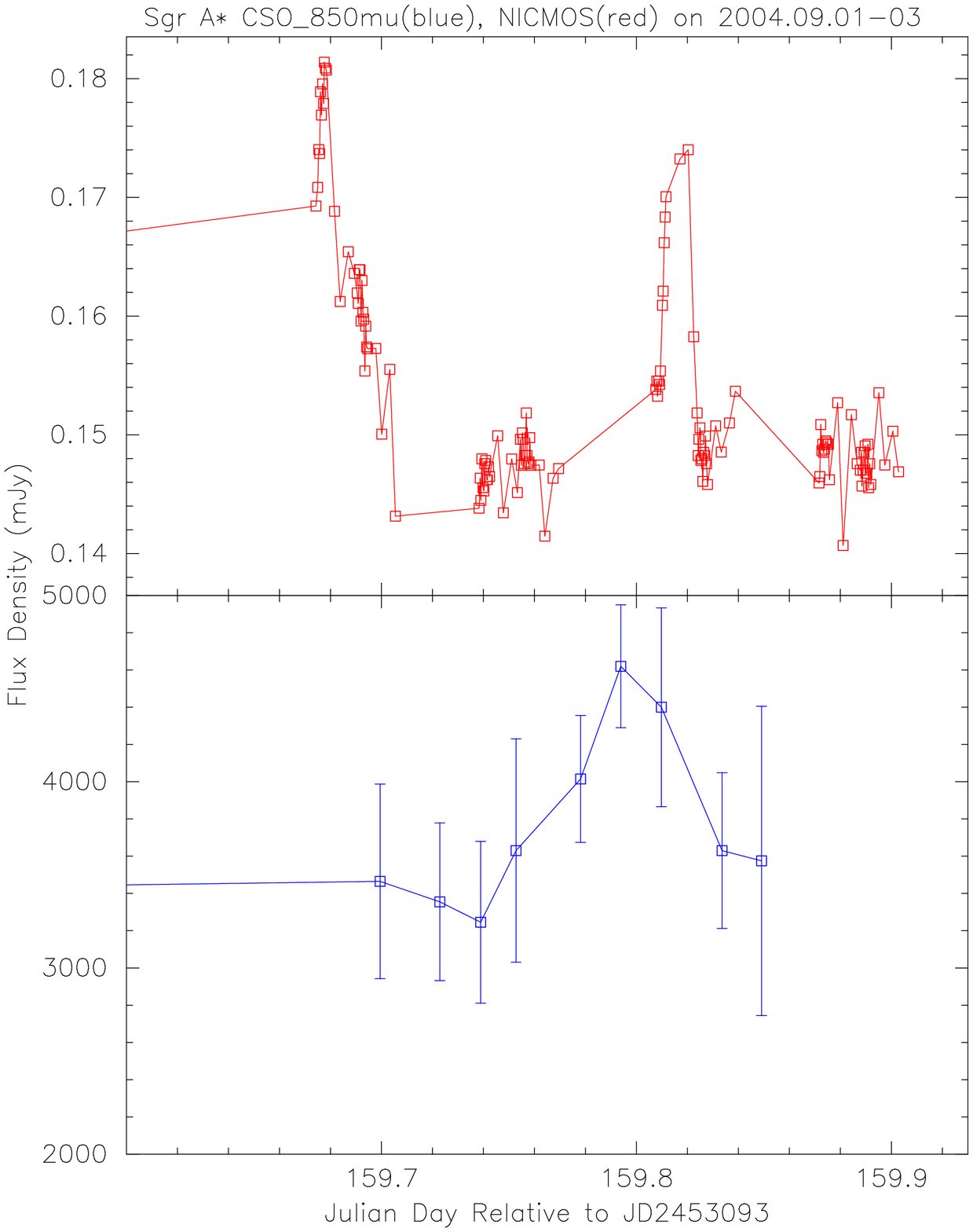}
\caption{(a -- Left)
The simultaneous near-IR
and X-ray  light curves using HST (top) and XMM (bottom), respectively. 
(b - Right)
The simultaneous near-IR (top)
and 850$\mu$m (bottom)  light curves using HST and CSO, respectively. 
}
\end{figure}

\subsection{X-ray and Near-IR Flares}

The long temporal
coverage of XMM-Newton and HST observations have led to the detection
of a simultaneous flare in both X-ray and near-IR wavelengths.  
Figure 1a shows the simultaneous
near-IR and X-ray emission with an amplitude increase of $\sim$15\% and
 100\% for the peak emission, respectively. 
It is clear that near-IR and X-ray flares track each other on 
short timescales, providing  compelling evidence for their correlation 
with no time delay. 
The X-ray light curve shows a double peaked maximum
flare which appears to be remarkably in phase with the
near-IR strongest double peaked flares, though with different amplitudes.
Since
detectable X-ray flares occur on the average once a day, 
the lack of X-ray counterparts
to other near-IR flares indicates  that i)
not all near-IR flares 
have 
X-ray counterparts or ii) that the X-ray-to-near-IR  ratio of some 
of the flares is  too small for X-rays to be detected above the strong, diffuse 
and steady X-ray background emission from the central parsec.  This 
observational fact has important implications on 
the emission  
mechanism, as described below.

We argue that the X-ray counterparts to the near-IR flares 
are unlikely to be produced 
by synchrotron radiation in the typical $\sim10$\,G magnetic field
 for two reasons.  First,
emission at 10\,keV would be produced by 100\,GeV electrons, which
have a synchrotron loss time of only 20\,seconds, whereas individual
X-ray flares rise and decay on much longer time scales (Baganoff et al. 2001).  
Second, the
observed spectral index of the X-ray flares, $S_\nu
\propto \nu^{-0.6}$ (Belanger et al. 2005), does not match the value of near-IR to 
X-ray
spectral index (Yusef-Zadeh et al. 2006a).
 We favor  an inverse Compton model
for the X-ray emission,
which naturally produces a strong correlation with the near-IR flares.
In this picture, sub-millimeter photons are up-scattered to X-ray  
energies by the electrons responsible for the near-IR synchrotron
radiation.  The fractional variability at sub-millimeter wavelengths  
is less than 20\%, so we consider 
a steady supply of submillimeter      
photons scattering off the variable population of GeV electrons that 
emit in the near-IR wavelengths.
In the ICS picture, the spectral index of the near-IR flare 
must match that of the X-ray
counterpart, i.e. $\alpha$ = 0.6. Future simultaneous observations 
should be able to test this picture.

Another important characteristic of the X-ray and near-IR flares is the
lack of a one-to-one match between them. 
This suggests that the variable spectral index (or the magnetic 
field) of 
near-IR flare emission  can  increase  the chance 
of detecting an X-ray flare counterpart if 
the energy spectrum of near-IR emitting particles  is 
soft.   
Support for the importance of the energy spectrum 
of the particles may come from 
 the sub-flares  in  the X-ray 
and near-IR flare light curves  showing   different 
amplitudes but the same phase (see Fig. 1a). 
Additional support for this suggestion comes from the 
fact that the brightest detected  X-ray flare thus far has shown the 
softest spectrum in X-rays (Porquet et al. 2003). 
Assuming that the ICS picture to produce X-ray emission from Sgr A* is 
correct, then 
it is  clear  that this mechanism 
is  vastly different than what is considered to produce X-ray emission from  micro-quasars.

\begin{figure}  
\includegraphics[width=0.45\textwidth]{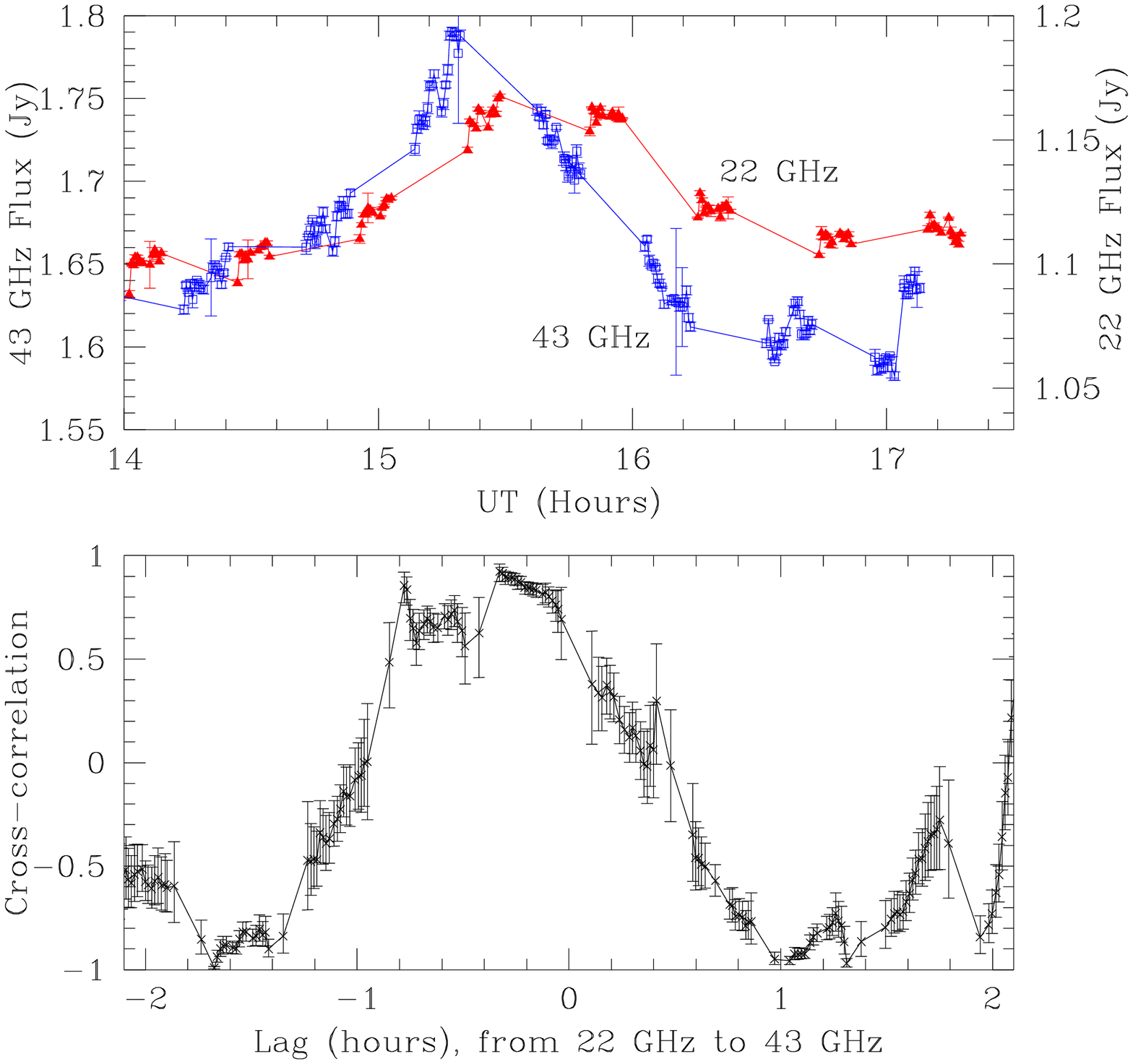}
\includegraphics[width=0.5\textwidth]{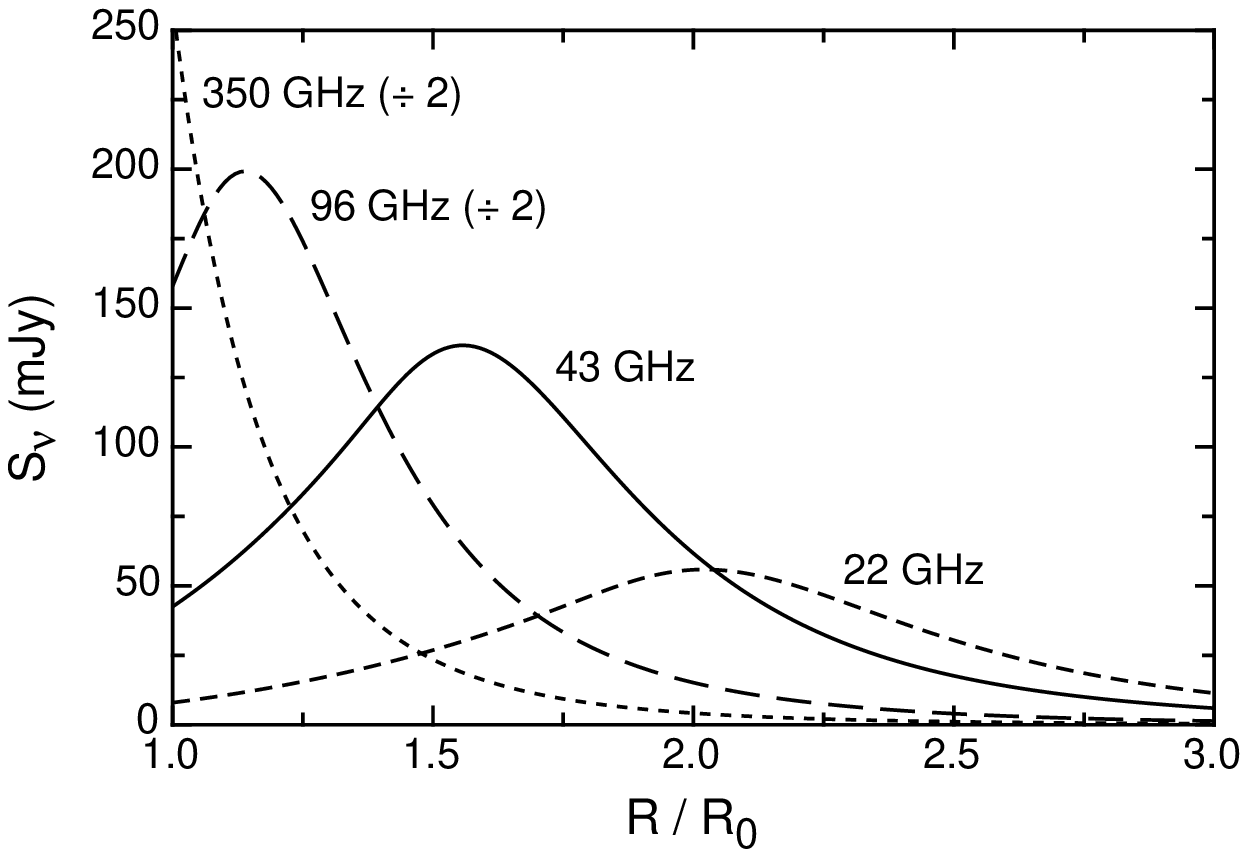}
\caption{(a - Left) The light curve of Sgr A* at 43 and 22 GHz with a
30s sampling
time (top) and the corresponding cross-correlation amplitude (bottom) as a function of
time. (b - Right) 
Synchrotron light curves at four different frequencies
for an
expanding blob of plasma with a peak near-IR flux of 1 mJy and R=4R$_s$.
}
\end{figure}

\subsection{Sub-millimeter and Near-IR Flares}

Figure 1b shows simultaneous near-IR and sub-millimeter
 observations using the NICMOS/HST and 
the CSO, respectively. Two near-IR flares with 
durations of about 20--35 min 
are displayed against an 850 $\mu$m flare with an estimated 2-hour 
duration.  Two important points can be 
drawn from this diagram. 
First, 
the simultaneity of the flares suggests  
that the same synchrotron emitting electrons are responsible for production of 
both 
near-IR and sub-millimeter flares. Thus, the duration of flares in these 
wavelength bands should be set by their cooling time scales. 
However the 2h time scale of the 850$\mu$m flare is inconsistent with
the expected 12h synchrotron cooling time scale. This implies that 
 the durations of the sub-millimeter and near-IR
flares must be set by dynamical mechanisms such as adiabatic expansion. 


The second point is that it is not clear whether the
sub-millimeter flare  is correlated simultaneously with the second bright
near-IR flare, or is produced by the first near-IR flare but with a delay 
of roughly 160 minutes (Figure 1b).  A delay 
between near-IR and sub-millimeter peak emission would imply that the 
emission is optically thick and cools by  
adiabatic  expansion of hot plasma. 

\begin{figure}
\includegraphics[width=0.475\textwidth]{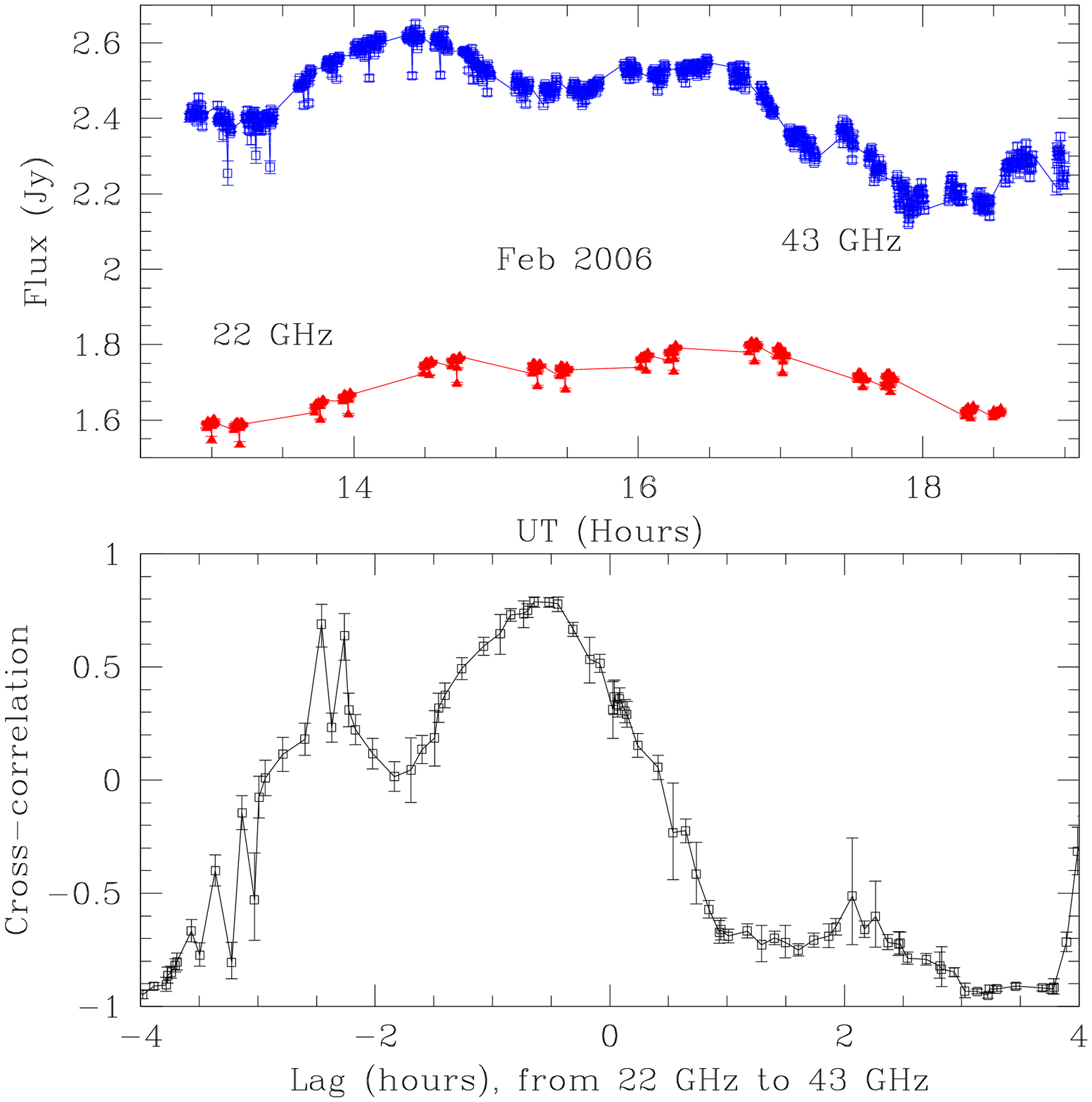}
\includegraphics[width=0.475\textwidth]{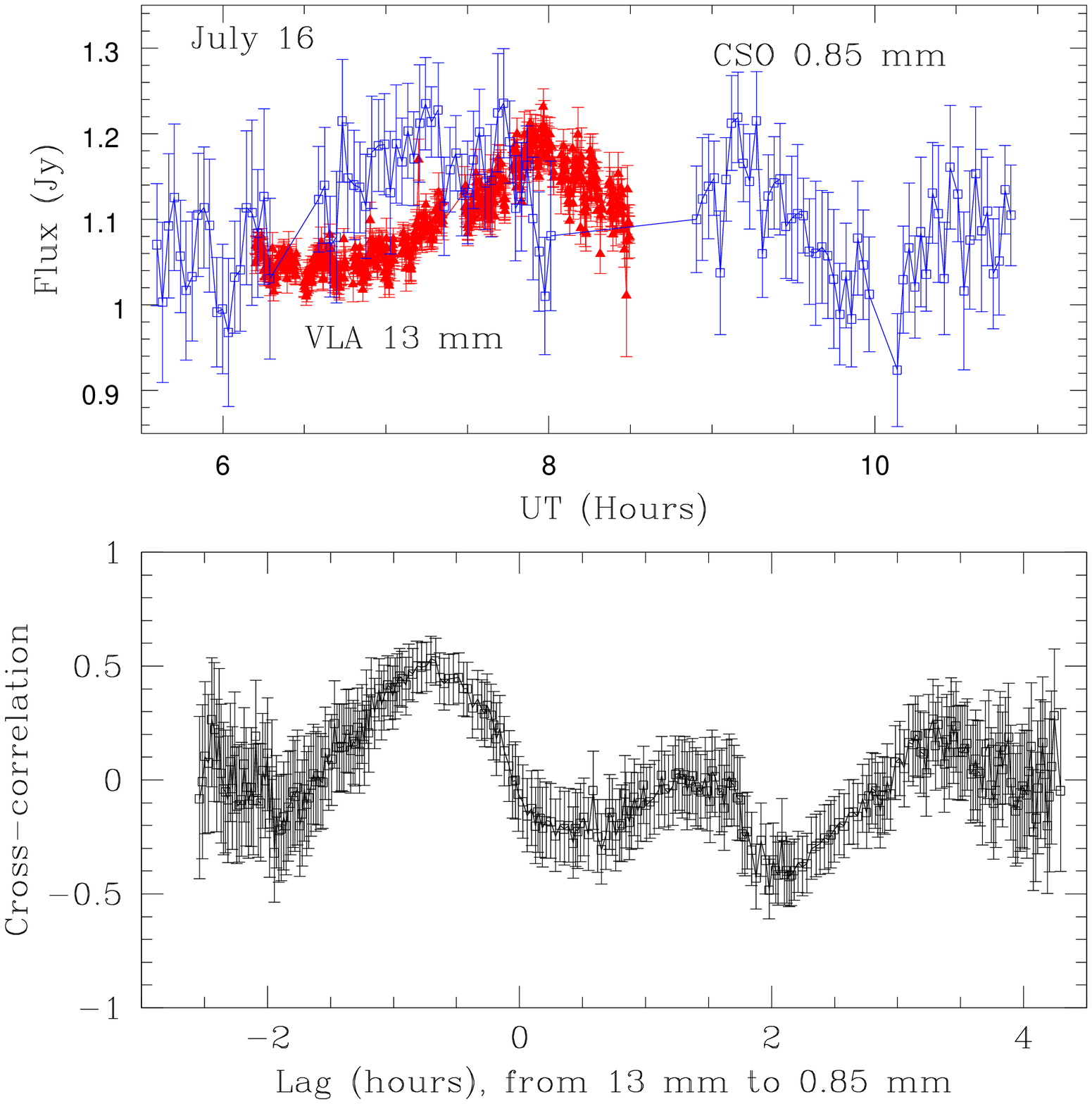}
\caption{ {(a - Left)}
The light curves of emission at 7mm (43 GHz) and 13mm (22 GHz) and the   
corresponding cross-correlation amplitude with a 9s sampling time (bottom). (b - Right)  Similar to (a) 
except that the observed light curves are measured at 13mm (23 GHz)  and 850 $\mu$m (350 GHz). 
}
\end{figure}  

\subsection{Radio  and Sub-millimeter Flares}

The durations of near-IR and sub-millimeter flares imply that flare activity can be 
driven by an outflow (Yusef-Zadeh et al. 2006a). This prediction motivated us to 
search for a time delay at radio wavelengths. 
In order to test the outflow picture, 
we studied the light curves of Sgr A* at  7 and 13mm  taken with the VLA 
in 
2005 (Fig. 2a).  
The light curves at both wavelengths show a 5-10\% increase of flux with respect to its 
quiescent level. 
We found that the emission at 13 mm lags that at 7 mm by $\sim$ 20 minutes
(Yusef-Zadeh et al. 2006b).  
If this time delay 
is correct,  then this  supports a picture in which the peak frequency of 
emission (i.e. near-IR) shifts toward lower frequencies 
(sub-millimeter, millimeter and then radio) as a self-absorbed synchrotron source 
expands adiabatically (van der Laan 1966). Figure 2b shows theoretical modeling of 
radio, millimeter, sub-millimeter and near-IR 
emission 
as a function of the size  of an expanding synchrotron bubble. 
For this plot, we used 
a particle energy spectrum $\propto E^{-3}$ and an initial blob size of 4R$_s$, where R$_s$ is the Schwarzschild radius. 
Simultaneous measurements were taken only at 7 and 13mm 
as the near-IR optically thin flare emission, with a flux density of 1 mJy,  is expected to arise from the most compact 
region. 
As the hot plasma expands, however, the larger surface area should produce optically thick millimeter and radio emission with a peak flux of 100--220 mJy, corresponding to a two to three order of magnitude increase in the flux of the optically thin emission.
This simple picture of adiabatic expansion of plasma is a phenomenon that is commonly observed  in 
micro-quasars (e.g. Ueda et al. 2002). 

Since synchrotron 
optical depth is $\propto\nu^{-2.5}$, an assumption that went into constructing Figure 2b is that the sub-millimeter 
peak emission 
is optically thin. However, this may not be the case, as Figure 1b
 alludes to a possible time delay 
between the near-IR and sub-millimeter peak emissions. 
Such a time delay 
should be longer 
for a flare with a steep spectral index 
than for one with a flat spectral index.  This is due to the larger number of 
steep spectrum nonthermal particles at low energies that can push the gas outward, producing a greater emitting surface. 
The hot material with a steep spectrum can also  keep the expanding plasma 
optically thick for a longer period, thus generating  a longer time delay. 


\begin{figure}
\includegraphics[width=0.7\textwidth]{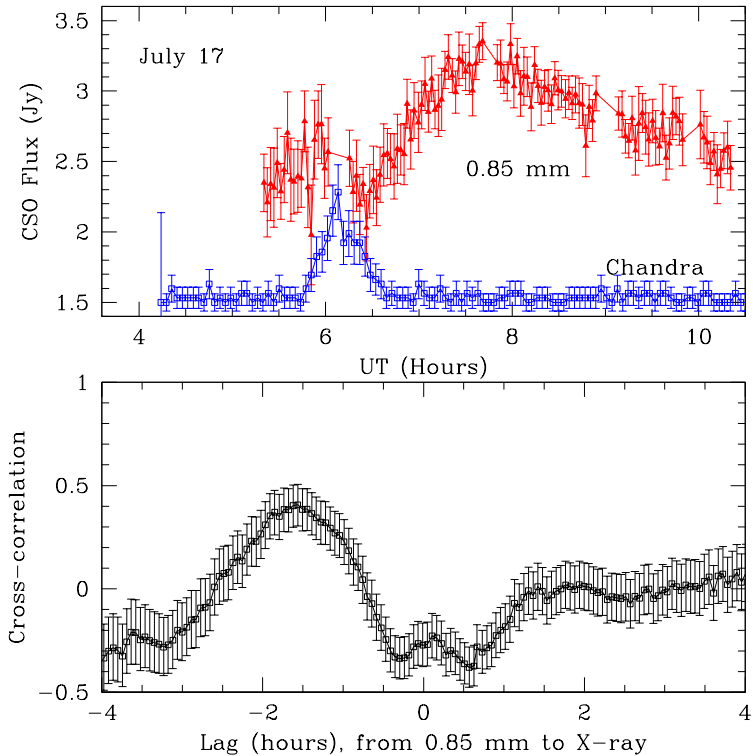}
\caption{ {}
The light curves of emission at 850 $\mu$m  and X-rays  (2-8 keV) and the   
corresponding cross-correlation amplitude. 
}
\end{figure}

To confirm the time delay between radio wavelengths, we recently re-observed 
Sgr A* with the VLA at 7 and 13mm at  two epochs in February and 
July of 2006. Figure 3a  shows an example of the light curves of Sgr A* at these 
wavelengths and the corresponding cross-correlation that shows  a peak which 
is consistent with the 7mm  emission leading the 13mm  emission. 
Figure 3b shows also a cross-correlation peak between 0.85mm and 13mm 
emissions using the CSO and the VLA. It can be seen that the 0.85mm emission 
leads the 13mm emission, consistent with our radio measurements.

\subsection{X-ray and Sub-millimeter Flares}

In the latest multi-wavelength campaign to study the flaring activity of 
Sgr A*, 
a flare was simultaneously detected at X-ray and sub-millimeter wavelengths. 
The 
results of this campaign will be given elsewhere. 
Here we briefly present the light curves of Sgr A* at 850$\mu$m and 
X-rays and the corresponding 
cross correlation peak, as displayed in 
Figure 4. 
The photon index of the 
X-ray flare is flat 1.0 (0.0, 2.6). The 90\% confidence error bars are large because the flare spectrum 
only had about 180 net counts.  The 
absorption corrected peak flux (2-8 keV) is 5.3$\times10^{-12}$ erg s$^{-1}$ cm$^{2}$.  Assuming 8 kpc as the 
distance to the 
Galactic center, the corresponding luminosity would be 4.0$\times10^{34}$ erg s$^{-1}$.
Although the sub-millimeter time coverage is not as good as the X-ray time coverage, we believe that 
the 
two peaks 
are related to each other. If so, the cross  correlation plot 
suggests that the sub-millimeter peak lags the X-ray peak by about 110 minutes.      
Another possible  support for the association of the two peaks comes from the flat shape of the decaying 
part of the two light curves in X-rays and sub-millimeter wavelengths, although the X-ray  flare decays somewhat faster than 
that of the sub-millimeter flare (see Morris et al. in this volume for the 
near-IR counterpart to this flare). Also, 
 assuming that the X-ray flare has a simultaneous  near-IR counterpart, 
then the $\sim1.5$ hour time delay seen in Figure 4 is consistent with that observed 
between the near-IR and sub-millimeter peaks in September 2004 (Yusef-Zadeh et al. 2006a).

\section{To Outflow or Not To Outflow}

Given the relatively short observing sessions for a southern source that 
is active most of the time, it is difficult to
definitively establish the relation between activity in various wavelength 
bands. However, the picture of flaring activity driving
an adiabatic expansion of hot plasma is consistent with observations, 
though it is not clear if this hot plasma can indeed leave
the gravitational potential of Sgr A*. As has been shown theoretically in 
the context of a jet model (see Melia and Falcke 2001;
Falcke \& Markoff 2000), the outflowing material requires a steep energy 
spectrum, a high initial velocity, a relatively large
initial size and a high fraction of thermal material mixed in with the 
nonthermal plasma in order for the initial hot blob to
escape with significant mass outflow rate. The 2-h duration of a typical 
radio flare corresponds to a maximum size scale of
$\sim15$ AU or 190R$_{s}$, assuming that the hot blob is expanding at the 
speed of light. The blob size at the peak is $\sim$100
R$_s$ which is is much bigger than than the VLBI size scale of Sgr A*. 
On 
the other hand, combining the 10$^3$s time delay and a
size scale of $<$ 1AU as the initial size of an expanding blob gives an 
expansion speed less than one-half the speed of light. This
estimate does not violate the inverse Compton limit and is consistent with 
the model described here. If the blobs originate near
the last stable orbit, it is not clear that this expanding material can 
leave the gravitational potential of Sgr A*. If Sgr A* has a
disk, the 20-40 minute time scale for the durations of near-IR flares can 
then be identified with the accretion disk's orbital
period at the location of the emission region. The dynamical time scale is 
the natural expansion time scale for the pressure to
build up, implying that the emitting size of the blob is of the order of 
the disk radius, a few R$_s$. In this picture, the hot
blobs expand to a size scale which is comparable to the 1 AU radio 
emission size of Sgr A*.


\begin{figure}
\includegraphics[width=0.5\textwidth]{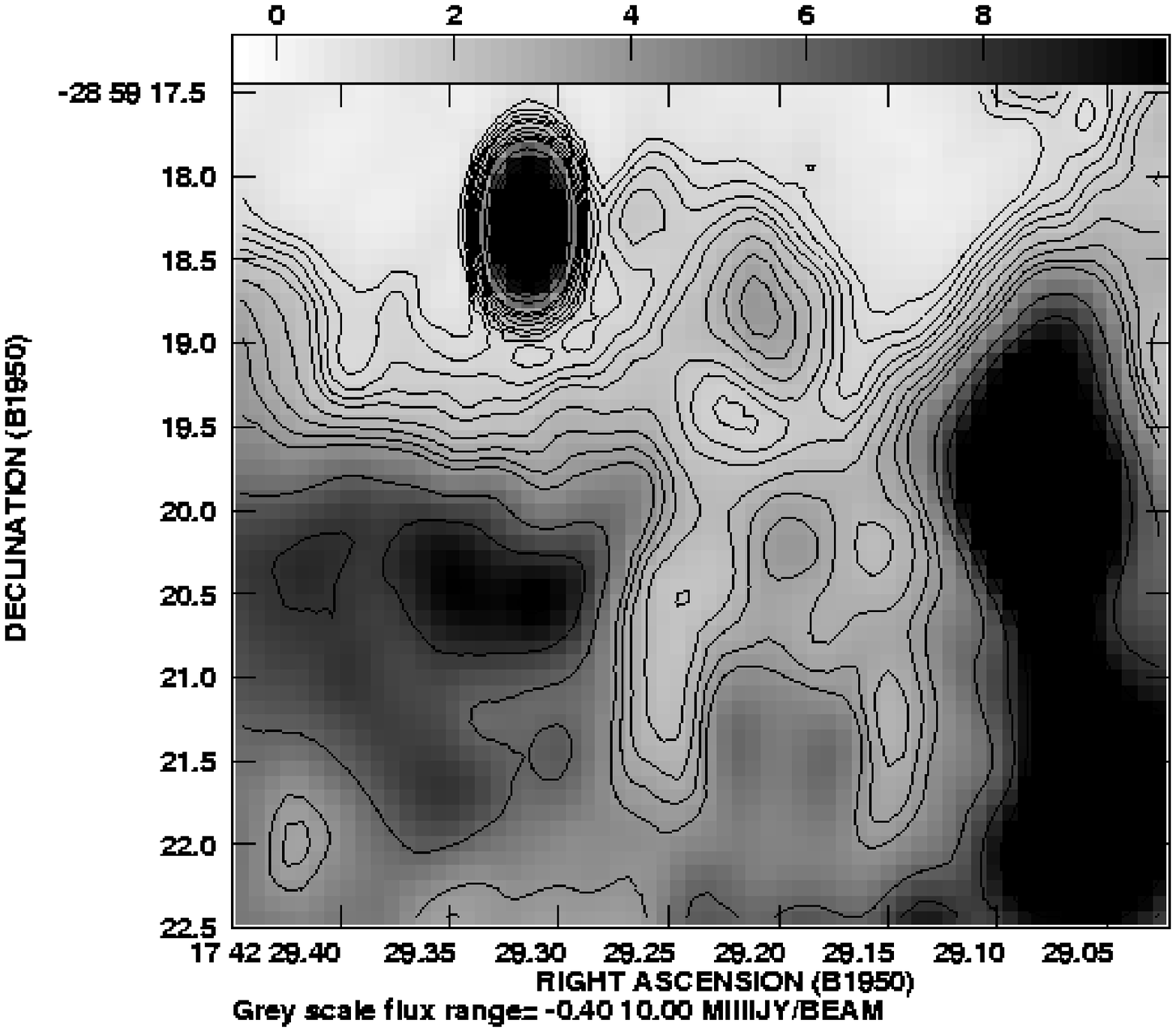}
\includegraphics[width=0.5\textwidth]{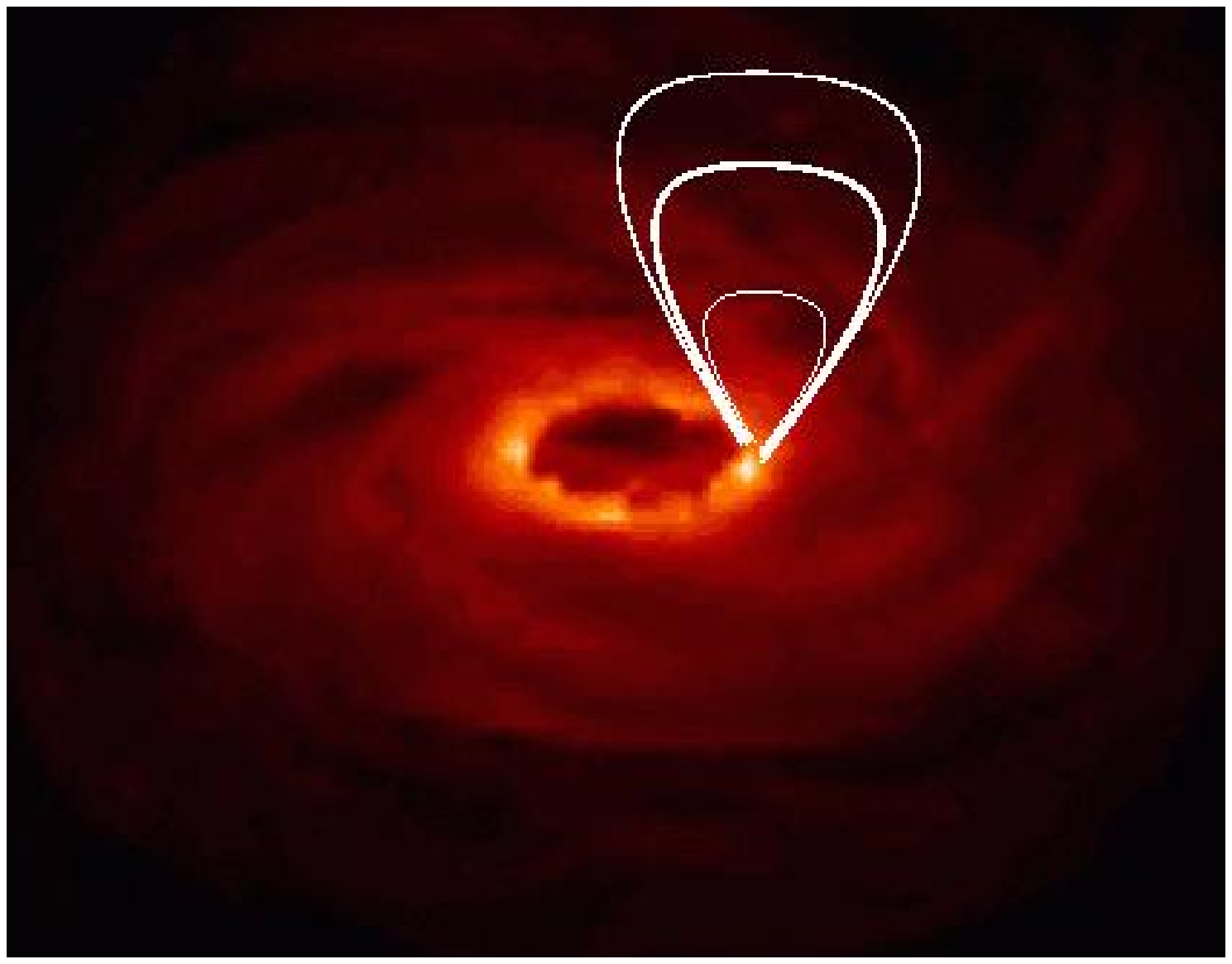}
\caption{ (a - Left) Contours of grayscale radio continuum emission from the inner
few arcseconds of Sgr A*. The dark compact source to the upper left 
coincides with Sgr A*. 
A number of ionized blobs with no apparent stellar counterparts have size  scales of about 10$^{17}$ cm 
and are 
apparent to the west 
 (right) of Sgr A* connecting  to the extended orbiting gas,  known as the mini-cavity. 
(b - Right) A schematic diagram of blobs expanding away from a disk  is 
superimposed on a simulated image
of the synchrotron emission from the central  16R$_s$ 
taken from Figure 2 of Goldston, Quataert \& Igumenshchev (2006). 
The brightest area of this disk model
comes from the region near the last stable orbit. 
}
\end{figure}  

It is also not clear if the outflowing material will be collimated since there is no evidence that 
Sgr A* has a disk. 
There are some hints that a collimated outflow may arise  from Sgr A*. 
For example, there  is the intriguing radio continuum image 
of the inner few arcseconds of Sgr A*,  as shown in 
Figure 5a (Yusef-Zadeh, Morris \& Ekers 1992). 
These large-scale
(10$^{17}$ cm) blob-like structures could potentially be 
material outflowing  from Sgr A*.  
One could speculate  that these blobs are collimated by a disk as shown in Figure 5b.
This figure shows a schematic diagram of expanding blobs 
superimposed on a  color image of 
 a synchrotron emitting disk 
 based on an MHD simulation of a  radiatively inefficient accretion flow 
(Goldston, Quataert \& Igumenshchev 2005).  
The observed time delay in radio wavelengths could also  be 
explained by a jet model (see Yusef-Zadeh et al. 2006b). 
In addition, one could argue that the similarity of the underlying 
ejection of hot plasma in Sgr A* and micro-quasars 
suggests that there should be a collimated outflow from Sgr A*.




\end{document}